\documentclass[11pt,a4paper]{article}
\usepackage[utf8]{inputenc}
\usepackage[english]{babel}
\usepackage{amsmath}
\usepackage{amsfonts}
\usepackage{amssymb}
\usepackage{lmodern}
\usepackage{booktabs}
\usepackage{pdfpages}

\newcommand{\err}{$\varphi$}

\newcommand{\errG}[1]{$\varphi (#1)$}

\newcommand{\cm}{$c$}
  
\newcommand{\reffigure}[1]{Figure~\ref{#1}}
\newcommand{\reftable}[1]{Table~\ref{#1}}
\newcommand{\refsection}[1]{Section~\ref{#1}}
\newcommand{\refequation}[1]{Equation~\ref{#1}}

\newcommand{\shortcite}[1]{\cite{#1}}

\begin{document}

\title{Estimating the sensitivity of centrality measures w.r.t. measurement errors}
\author{Christoph Martin \and Peter Niemeyer}
\date{%
Leuphana University of L{\"u}neburg \\
 Universitätsallee 1, 
         21335 L{\"u}ne­burg, Germany \\
\texttt{\{cmartin,niemeyer\}@uni.leuphana.de} \\
\vspace{0.5cm}
    September 20, 2017
}

\maketitle

\label{firstpage}

\begin{abstract}

Most network studies rely on an observed network that differs from the underlying network which is obfuscated by measurement errors. It is well known that such errors can have a severe impact on the reliability of network metrics, especially on centrality measures: a more central node in the observed network might be less central in the underlying network.

We introduce a metric for the reliability of centrality measures -- called sensitivity.
Given two randomly chosen nodes, the sensitivity means the probability that the more central node in the observed network is also more central in the underlying network.
The sensitivity concept relies on the underlying network which is usually not accessible.
Therefore, we propose two methods to approximate the sensitivity.
The iterative method, which simulates possible underlying networks for the estimation and the imputation method, which uses the sensitivity of the observed network for the estimation.
Both methods rely on the observed network and assumptions about the underlying type of measurement error (e.g., the percentage of missing edges or nodes).

Our experiments on real-world networks and random graphs show that the iterative method performs well in many cases. In contrast, the imputation method does not yield useful estimations for networks other than Erdős–Rényi graphs. 

\end{abstract}


\section{Introduction}
Measurement errors in network data are a central problem in the field of network analysis.
Virtually all empirical network data is affected by a certain kind of measurement error and previous research has shown that these errors often have a major impact on the results of network analysis methods, especially on centrality measures ~\cite{Wang2012}.
For example, a more central node in the observed (erroneous) network might be less central in the hidden (unobserved, error-free) network.
To quantify the impact of measurement errors on centrality measures, we introduce a metric -- called sensitivity.
Given two randomly chosen nodes, the sensitivity means the probability that the more central node in the observed network is also more central in the underlying network.

Currently, most applied network studies only report that measurement errors might have affected the data collection (e.g., due to absence of actors at the day of the survey or due to the study design). Most of the time, however, the impact of these measurement errors on centrality measures are not discussed. This might be due to the fact that there is currently no established way to estimate the sensitivity for an observed network.

Some researchers have investigated the impact that different kinds of measurement errors 
have on the reliability of centrality measures in the case of random graphs and real-world networks~\cite{Costenbader2003, Borgatti2006a, Kim2007, Frantz2009a, Wang2012, Smith2013, Platig2013, Niu2015, Lee2015}.
For an extensive survey of previous studies about the reliability of centrality measures see Smith et al. (2017) \cite{Smith2017}.
These important studies provide guidelines for researchers on how to design future studies (e.g., what kind of measurement error might be especially harmful in a given scenario) and suggestions on which centrality measure might be more reliable in a given scenario.
Unfortunately, it is difficult to identify general patterns for the reliability of centrality measures in real-world networks.
As common sense suggests, centrality measures become less reliable with an increasing level of error.
Additionally, the particular relationship between error level and reliability is highly dependent on the type of measurement error, the centrality measure, and the network structure.

There are studies that use the observed network to reconstruct the hidden network, estimate statistics about the hidden network, or estimate the influence that measurement errors have on network analysis methods 
\cite{Butts2003, Huisman2009, Handcock2010, Kim2011, Frantz2016,
Wang2016, Newman2017}.
Despite their important contributions, these studies usually focus on network invariants (e.g., diameter or average path length) and do not explicitly address node invariants (e.g., the centrality values for all nodes in a network).

Studies about the robustness of networks, the capacity of networks to maintain functionality when the network is modified,  also usually focus on network invariants. In this paper, however, we do not study the robustness of networks. For an extensive overview about the robustness of networks see Barabási (2016) \cite{Barabasi2016} and Havlin \& Cohen (2010) \cite{Havlin2010}.

The sensitivity concept relies on the hidden network which is usually not accessible.
In this paper, we propose two methods (``imputation method" and ``iterative method") that allow the researcher to estimate the sensitivity of the hidden network, given the observed network and some assumptions about the measurement error.
In case of the imputation method, we try to simulate the hidden network by inverting the measurement error (e.g.: if 10\% of randomly chosen edges are missing, we add 11,11\% edges by linking randomly chosen pairs of non-adjacent nodes) and then compute the sensitivity with respect to the simulated hidden network.
In case of the iterative method, we assume that the considered hidden network has a certain kind of self-similarity property. That is, we assume that the sensitivity of the observed network (regarding the assumed measurement error) is a good estimate of the sensitivity of the hidden network.
While it is known that imputation techniques may work well \cite{Rubin1996,Allison2002, Schafer2002}, simulating appropriate hidden networks is complicated and calculating the estimate based on the imputation method is, therefore, difficult.
In contrast, the estimate based on the iterative method is relatively easy to calculate. However, this method relies on a self-similarity property that is difficult to verify.
In addition to the estimation methods, we introduce an easy-to-interpret measure for the reliability of centrality measures and a generic concept to model measurement errors.
We test both estimation methods on random graphs and real-world networks.
While the imputation method and the iterative method perform well
on Erdős–Rényi graphs (ER~graphs), the iterative method, which is easier to calculate, performs much better in case of Barabási–Albert~graphs~(BA~graphs) and real-world networks. For real-world networks, the iterative method performs especially well in case of the PageRank.
To measure the reliability of centrality measures, we introduce the necessary concepts in \refsection{basicconcepts}. The estimation methods are presented in \refsection{sec:estimateDesc}. In \refsection{sec:applicationToRw}, we apply these methods to real-world networks. Results are discussed in \refsection{sec:discussion}.

\section{Basic concepts}
\label{basicconcepts}
Let G be an undirected, unweighted, finite graph with vertex set $V(G)$ and edge set $E(G)$.\footnote{In 
this study,  we consider unweighted and undirected graphs.
However, most of our concepts  can be extended to directed and weighted graphs.}
A centrality measure \cm\ is a real-valued function that assigns centrality values to all nodes in a 
graph and is invariant to structure-preserving mappings, i.e., centrality values depend solely on the structure of a graph. External information (e.g., node or edge attributes) have no influence on the centrality values (cf. Koschützki et al. (2005) \cite{Koschutzki2005}).
We denote the centrality value for node $u \in V(G)$ by $c_G(u)$ and the centrality values for all 
nodes in G~$(u_1, u_2, \dots, u_n)$ by the vector $c(G) := (c(u_1), \dots, c(u_n))$. 

\label{sec:DescriptionCentralityMeasures}
The following centrality measures are used in this study: closeness centrality, betweenness centrality \cite{Freeman1978}, degree centrality, eigenvector centrality \cite{Bonacich1987}, and the PageRank \cite{Brin1998}. All centrality measures are calculated using the igraph library (version 0.7.1, Csardi \& Nepusz~\shortcite{Csardi2006}).

Let G and $G'$ be two graphs and \cm\ a centrality measure.
A pair of nodes $u,v \in V(G) \cap V(G')$ and $u\neq v$  is called concordant w.r.t. $c$
if both nodes have distinct centrality values and the 
order of u and v is the same in c(G) and c($G'$), i.e., either $c_G(u) < c_G(v)$ and $ c_{G'}(u) < 
c_{G'}(v)$ or 
$c_G(u) > c_G(v)$ and $ c_{G'}(u) > c_{G'}(v)$.
A pair of nodes is called discordant 
if both nodes have distinct centrality values and the order of u and v in c(G) differs from the order of u and v 
in c($G'$), 
i.e., either $c_G(u) < c_G(v)$ and $ c_{G'}(u) > c_{G'}(v)$ or 
$c_G(u) > c_G(v)$ and $ c_{G'}(u) < c_{G'}(v)$.
Ties are neither concordant nor discordant.

A random graph consists of a finite set of graphs $\Omega$
equipped with a function $P$ that assigns a probability to
every graph in this set
(cf. Bollobás \& Riordan (2002) \cite{Bollobas2002}).

\subsection{Modeling measurement errors}
\label{sec:DescriptionErrorMechanisms}
Network data can be influenced by a variety of different measurement errors. Recently, 
Wang~et~al.~\shortcite{Wang2012} categorized measurement errors into six groups: false negative nodes and edges, 
false positive nodes and edges, and false aggregation and disaggregation.
For example, when 10\% of the 
edges are missing in the observed network data, the graph constructed from this observed data suffers from 
false negative edges.

To describe measurement errors, we introduce the notion of an error mechanism.
An error mechanism \err\ is a procedure that describes measurement errors which may occur during the data 
collection. For a graph G, \errG{G} is a random graph which probability distribution for the 
possible graphs depends on the error procedure that describes \err. Hence, all graphs that could be 
observed when the error procedure influences the data collections are possible outcomes of this 
random graph.

\begin{figure}[htbp]
\includegraphics[scale=.85]{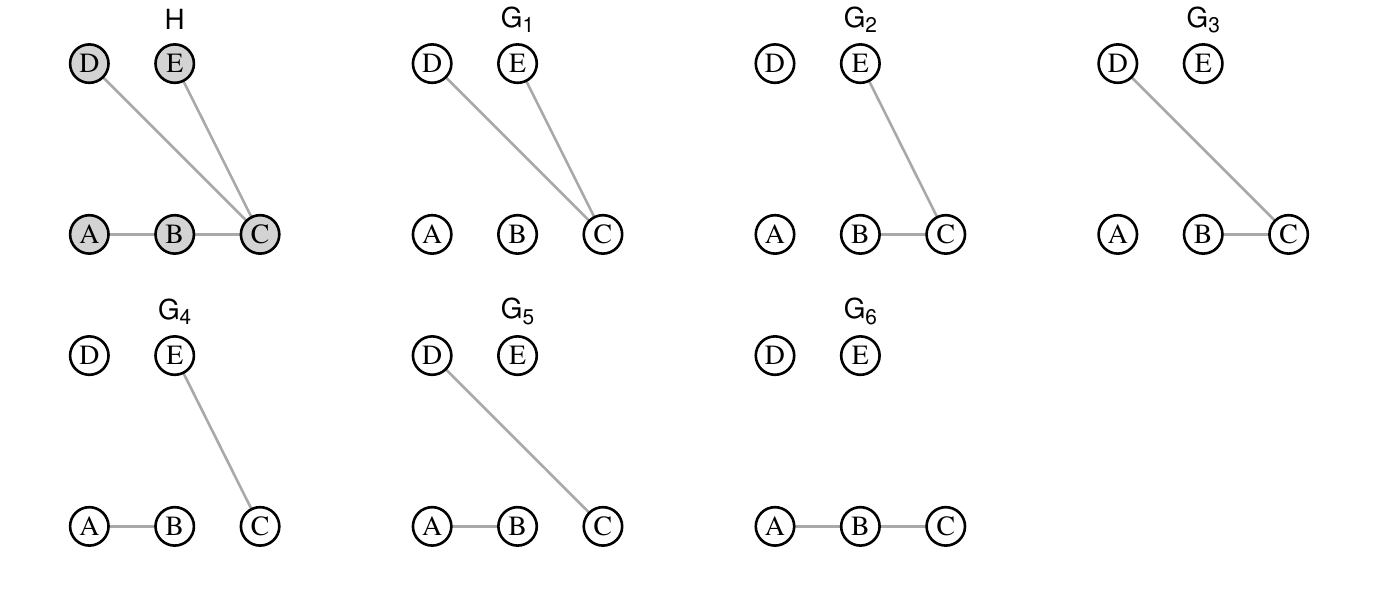}
\caption[Shortened figure caption for the list of illustrations]
{
In this example, \err\ is defined as the error mechanism ``50\% of all edges are missing 
uniformly at random". Hence, \errG{H} is a random graph with possible outcomes 
$\Omega = \{G_1, G_2, \dots, G_6\}$ and $P(G_i) = \frac{1}{6}$.
}
\label{fig:exampleError}
\end{figure}

To illustrate the concept of an error mechanism, consider the graphs illustrated in 
Figure~\ref{fig:exampleError}. The initial graph is denoted by H (drawn in the upper left corner).
We assume that we know the error mechanism that compromises the data collection.
For this example, we assume that the error mechanism 
\err\ is
edges missing uniformly at random with an error level of 50\%.
All graphs in the set of possible outcomes for this random graph $\Omega = \{G_1, G_2, \dots, 
G_6\}$ are also shown in \reffigure{fig:exampleError}.
In this example, the probability function is
$P(G_i) = \frac{1}{6}$; all graphs in $\Omega$ occur with the same probability. 
However, this concept is not limited to a uniform distribution.

In general, error mechanisms can rely on node or edge attributes. In this study, we focus on four common error mechanism that do not depend on external attributes:

\begin{enumerate}
\item
Nodes missing uniformly at random (rm nodes): A fraction of nodes (and all edges connected to these nodes) is missing in the observed network. All nodes have the same probability to be missing in the observed network.

\item
Edges missing uniformly at random (rm edges unif.): A fraction of edges is missing in the observed network. All edges have the same probability to be missing in the observed network.

\item
Edges missing proportional (rm edges prop.): A fraction of edges is missing in the observed network. The probability that an edge is missing in the observed network is proportional to the sum of the degree values of the endpoints.

\item
Spurious edges (add edges): The observed network contains too many edges. Every non-existing edge has the same probability to be erroneously observed.
	\end{enumerate}

Imputation techniques are commonly used to replace missing data with plausible estimates. Inspired by Huisman~\shortcite{Huisman2009},
we define an imputation mechanism as a procedure that aims to ‘undo’ the effects that measurement errors have on network data.

\label{sec:ImputationCounterexample}
For a given error scenario, there might be multiple or no appropriate imputation mechanisms. 
Moreover, the choice of the appropriate imputation mechanism depends on the type of measurement 
error and possibly also on the network structure.
For example, if edges are missing uniformly at random, a possible imputation mechanism is to add 
edges randomly (uniform) to the observed network. However, the success of this procedure depends on 
the structure of the network. Consider the scenario that the hidden network is a star graph. In this case, it is quite likely 
that this imputation approach would have a negative impact on subsequent analyses because it is much 
more likely that the imputed edges are between leaf nodes than between the internal node and a leaf 
node. Hence, if the structure of the hidden network is not known (which is usually the case), it is 
difficult to define an appropriate imputation procedure.

We use $\psi$ to denote a specific imputation mechanism.
Analogously to error mechanisms, $\psi(G)$ denotes a random graph. All graphs that can occur when the imputation mechanism $\psi$ is applied to graph $G$ are possible outcomes for this random graph.
In this paper, we use three generic imputation mechanisms:
\begin{enumerate}
\item
If $k$ edges are missing in the observed network, we choose $k$ edges uniformly at random  
from the set of edges that are not in the observed network and add them to the observed network.
\item
If there are $k$ spurious edges in the observed network, we choose $k$ edges that do exist in the observed network and delete them.
\item
If $k$ nodes are missing, we choose the degrees for $k$ new nodes from the current degree distribution and successively add these nodes to the networks. The links between new and existing nodes are created uniformly at random.\footnote{
There are numerous possibilities how to connect the new nodes to the existing ones. For example, one could also use a preferential attachment like approach.
This variety of options illustrates once again, that choosing an appropriate imputation mechanisms is challenging.
}
\end{enumerate}

\subsection{Sensitivity of centrality measures}
\label{sec:sensitivityDefn}
%

Let G and $G'$ denote graphs on the same vertex set and \cm\ a centrality measure.
The sensitivity of the centrality measure \cm\ w.r.t. those two graphs is the probability that
two nodes with distinct centrality values, randomly chosen from the vertex set of G and $G'$, have the same order in c(G) and c($G'$), i.e., they are concordant.
If this quantity is close to one, the centrality measure is considered to be robust. If it is close to zero, the centrality measure is considered to be sensitive. We calculate the 
sensitivity
$\rho$ for a centrality measure \cm\ with respect to G and $G'$ as follows:

\begin{equation}
\rho_c(G, {G'}) = \frac{n_c}{n_c+n_d} 
\end{equation}

\noindent
With $n_c$ as the number of concordant pairs and $n_d$ as the number of discordant
pairs w.r.t. the order given by $c(G)$ and $c({G'})$.\footnote{
It may occur
that $V({G'}) \neq V(G)$. In these cases, we only consider entries in $c(G)$ and $c({G'})$ that
correspond to nodes that are in both graphs (G and $G'$). This is a common approach for the comparison 
of graphs on different vertex sets \cite{Wang2012}.
}

The sensitivity as defined above is closely related to Goodman and Kruskal's rank
correlation coefficient $\gamma$ which is the difference between concordant and discordant pairs
divided by the sum of concordant and discordant pairs \cite{Goodman1954}.
We can use this relationship to calculate the sensitivity as follows:

\begin{equation}
 \rho_c(G, {G'}) = \frac{\gamma (c(G), c({G'}))+ 1}{2} .
\end{equation}

Let us apply this concept to a graph illustrated in \reffigure{fig:exampleError}.
Assume that we have observed the graph labeled as $G_6$ and that we are interested in the sensitivity of  
the degree centrality.
Then, the degree centrality values are $deg(H) = (1,2,3,1,1)$ 
 and $deg(G_6) = (1,2,1,0,0)$. Based on the degree values, we can 
calculate the sensitivity of the degree centrality with respect to $G_6$ and H:
$\rho_{deg}(G_6,H) = \frac{5}{6}$. If we randomly choose a pair of nodes, there is a 0.83 chance that 
their order induced by the degree centrality is the same in both graphs.

\subsection{Exemplary application of the sensitivity concept}
\label{errobustness}
In this section, we apply the concepts introduced above to analyze the sensitivity of 
centrality measures in ER graphs.

For all combinations of centrality measures and error mechanisms  introduced in 
Section~\ref{sec:DescriptionCentralityMeasures} and \ref{sec:DescriptionErrorMechanisms}, we 
perform the experiment described below 500 times. For every error mechanism, we consider two cases, a moderate scenario of 10\% error level and a more intense scenario with 30\% error level. 
The procedure for the experiment is as follows:

\begin{enumerate}
\item
Generate an ER graph with 100 nodes and edge probability 0.2 and denote it by H. This is the error-free (hidden) graph which is not available to the researcher.
\item
Choose a graph from \errG{H} and denote it by O. This is the observed graph which is affected by measurement errors.
\item
Calculate the sensitivity $\rho_c(O, H)$.
\end{enumerate}

\begin{figure}[htbp]
\includegraphics[width=1\textwidth]{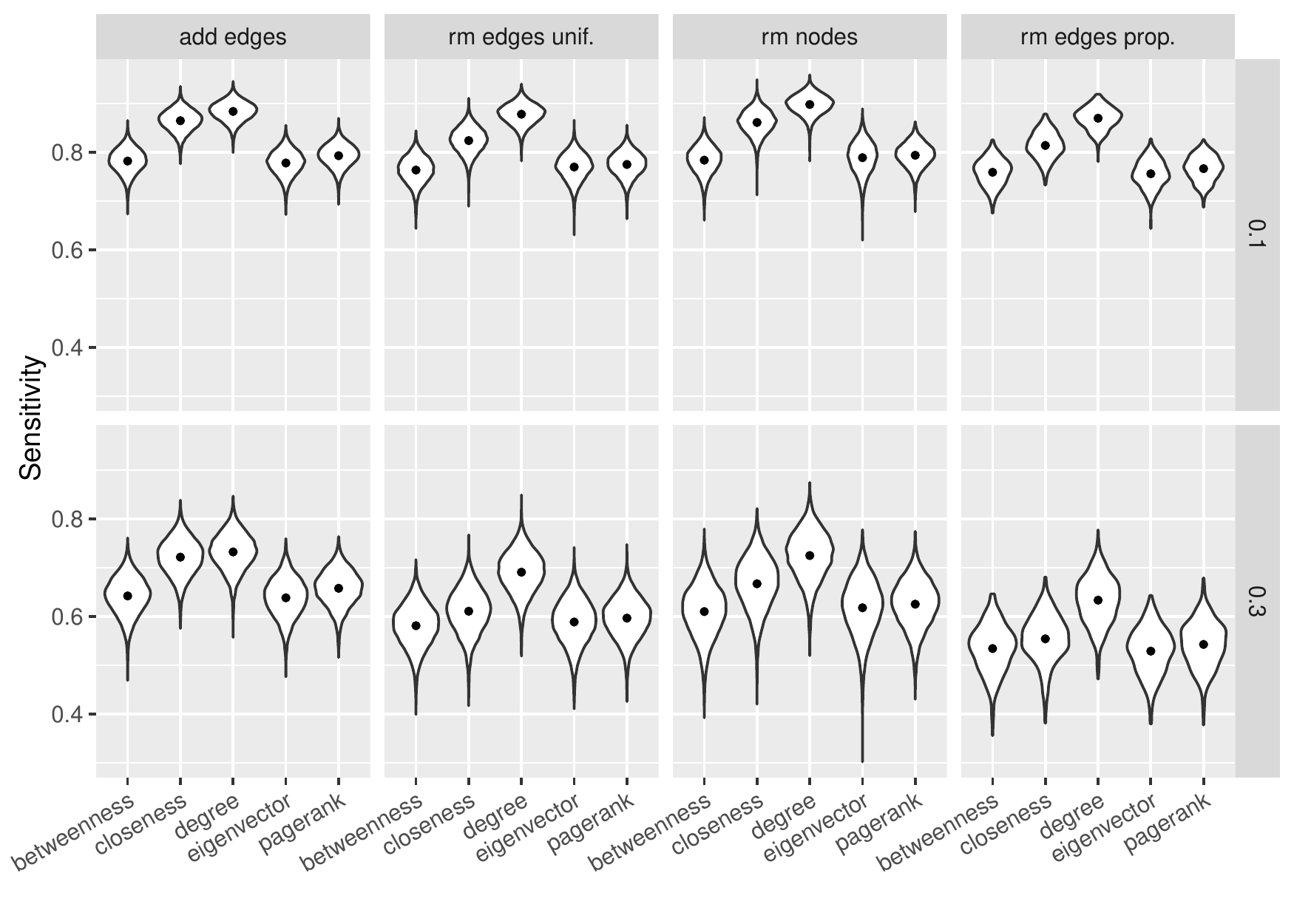}
\caption[Shortened figure caption for the list of illustrations]
{
Sensitivity of centrality measures in ER graphs. The violin plots in each panel show the distribution of the sensitivity of the corresponding centrality measure under the influence of different error mechanisms.
Mean values are indicated by a black dot.
}
\label{erexample}
\end{figure}

The results are shown in Figure~\ref{erexample}. Every panel shows violin plots for the distribution 
of the sensitivity of the centrality measures. Despite the fact that these graphs are very 
homogeneous, we make interesting observations. The sensitivity differs between centrality measures. 
The degree centrality is, in all cases, the most robust measure. Generally, the sensitivity values 
for the 30\% error mechanisms are much lower than for the 10\% error mechanism. The variance of the 
sensitivity values also increases with increasing error level.
Usually, the sensitivity also depends on the error mechanism. However, this is not always the case.  
(e.g., degree centrality the the case of 10\% error mechanisms). These observations are  conclusive 
with the results of  \cite{Borgatti2006a}. These results show that measurement errors have severe consequences for the reliability of centrality 
measures, even for homogeneous networks
such as ER graphs.

\section{How to estimate the sensitivity of a centrality measure}
\label{sec:estimateDesc}

In a lot of network studies, the observed network data contains sampling errors \cite{Leecaster2016, Schulz2016, Wang2016}.
But in general, the authors of such studies have no tools to describe the impact of sampling errors on the network measures (e.g., centrality measures) they apply.
In general, the assumptions made on sampling errors are mentioned in the limitations, but they are not considered as part of the network model.

The sensitivity concept as introduced in \refsection{sec:sensitivityDefn} helps researchers to describe this impact: given the observed network $O$, the (unknown) hidden network $H$, and a centrality measure $c$, $\rho_c(O, H)$ measures the probability, that two randomly chosen nodes (with distinct centrality values) have the same order in $c(O)$ and $c(H)$.
Hence, $\rho_c(O, H)$ is capable to measure the impact of the sampling error on the centrality measure $c$. Thus, the sensitivity can be used to measure the reliability of a centrality measure 
with respect to sampling errors. We call $\rho_c(O, H)$ the ``true sensitivity".

Unfortunately, the hidden network $H$ is not known and thus the true sensitivity cannot be computed explicitly.
In this section, we propose two methods for the estimation of the true sensitivity based on the observed network $O$. Moreover, we provide an example for their application and demonstrate how the estimation results can be evaluated.

\subsection{Methods to estimate the sensitivity}
\label{sec:MethodDescription}
In this section we propose two methods for the estimation of the true sensitivity $\rho_c(O, H)$.
In addition to the observed network $O$ and a centrality measure $c$, each of these
 methods needs an additional assumption.\footnote{
 Both methods are not limited to our definition of sensitivity. It would be interesting to see results for other metrics, for example,  the estimation of the most central node (see Frantz \& Carley (2016) \cite{Frantz2016}).}

The first method that we propose is the imputation estimate for the sensitivity of a given centrality measure (``imputation method").
Based on the observed network O and an imputation mechanism $\psi$,  we try to ``reconstruct" the hidden network from  the observed network. Based on the reconstructed network and the observed network, we calculate the estimate for the true sensitivity of the hidden network H and a centrality measure $c$ as follows:
\begin{equation}
 \hat{\rho}_c^{imp}(O, H) := E( \rho_c(O, \psi (O))).
\end{equation}
\noindent
Since $ \psi (O)$ is a random graph, $ \rho_c(O, \psi (O))$ is a random variable and
we use the expected value of this expression as the estimate for the sensitivity. 
The actual form of the imputation mechanism $\psi $ depends strongly on the error mechanism $\varphi$ which has influenced the data collection. However, depending on the network structure and the error mechanism, it might be very difficult to define an appropriate imputation mechanism (see \refsection{sec:ImputationCounterexample}).

The second method that we propose is the iterative estimate for the sensitivity of a given centrality measure (``iterative method"). If we assume that the network of interest has a self-similarity property in the sense that subgraphs of this network have the same sensitivity as the initial network, we can apply the (assumed) error mechanism \err\ to the observed network and calculate the estimate for the true sensitivity of the hidden network H and a centrality measure $c$ as follows:
\begin{equation}
 \hat{\rho}_c^{iter}(O, H) := E( \rho_c(O, \varphi (O))).
\end{equation}
\noindent
Since $ \varphi (O)$ is a random graph, $ \rho_c(O, \varphi (O))$ is a random variable and
we use the expected value of this expression as the estimate for the sensitivity.
Our experiments indicate that this self-similarity property may exist in many cases even though it is hard to prove that such a property does exist in complex networks.

Both methods do rely on assumptions that are difficult to prove (appropriate imputation mechanism and self-similarity property). However, if these assumptions hold, we should be able to make good estimates for the sensitivity.

\subsection{Example for method application}
\label{sec:ERexperimentSetup}
As a fist step to verify whether the proposed methods yield useful results,
we apply the four error mechanisms to ER graphs and try to predict the sensitivity using the imputation method as well as the iterative method.

For all combinations of centrality measures and error mechanisms  introduced in 
Section~\ref{sec:DescriptionCentralityMeasures} and \ref{sec:DescriptionErrorMechanisms}, we 
perform the experiment described below 500 times. For every error mechanism, we consider two cases, a moderate scenario of 10\% error level and a more intense scenario with 30\% error level. 
We perform the following steps to simulate erroneous data collection and to collect 3-tuples of true 
sensitivity, the estimate based on the iterative method, and  the estimate based on the imputation method:
\begin{enumerate}
\item
We generate an ER graph with 100 nodes and edge probability 0.2 and denote it by~$H$. This graph represents the (error-free)
hidden network.\footnote{
Our experiments have shown that the choice of $p$ has little influence on the main results associated with this section. Hence we will only consider the case of $p = 0.2$.
}
\item
We choose a graph from \errG{H} and denote it by $O$. This graph represents the observed network which is 
affected by measurement errors. For evaluation purposes, the true sensitivity  $\rho_{c}(H, O)$ is 
calculated and denoted by $s$.
\item
Based on the observed network $O$, two estimates for the true sensitivity are calculated.
The imputation estimate ($\hat{\rho}_c^{imp}(O, H) $) is denoted by $\hat{s}^{imp}$,
the estimate calculated according to the iterative method ($\hat{\rho}_c^{iter}(O, H) $) is denoted by $\hat{s}^{iter}$.
\end{enumerate}

\noindent
The results of these experiments are 
listed in \reftable{tab:er95res}.
Values in the $s$ columns represent the mean values of the true sensitivity for 
all 500 runs. The 95th percentile values of the absolute difference between the true sensitivity and the 
estimate are labeled with $\hat{e}^{imp}$ for the imputation estimates and $\hat{e}^{iter}$ for the iterative estimates. We call this value the absolute error. For example, the average true sensitivity of 
the betweenness centrality under the influence of the error mechanism add edges random 10\% is 0.891 and 
the imputation  estimate of the true sensitivity is in the interval [0.871, 0.911] in 95\% of all runs.


\begin{table}[h]
\centering
\caption{Results for the estimate of the sensitivity (ER~graphs). 
For all centrality measures, average true sensitivity (column s) and 95th percentile of the absolute error for the estimate ($\hat{e}^{imp}$: imputation method; $\hat{e}^{iter}$: iterative method) are shown. (Values are multiplied by 100 for better readability.) }
\label{tab:er95res}
\setlength{\tabcolsep}{0.05em} 

\begin{tabular}{lrrrrrrrrrrrrrrr}
\hline\hline
                & \multicolumn{3}{l}{betweenness}           & \multicolumn{3}{l}{closeness}             & \multicolumn{3}{l}{degree}                & \multicolumn{3}{l}{eigenvector}           & \multicolumn{3}{l}{PageRank}              \\
                & s    & $\hat{e}^{imp}$ & $\hat{e}^{iter}$ & s    & $\hat{e}^{imp}$ & $\hat{e}^{iter}$ & s    & $\hat{e}^{imp}$ & $\hat{e}^{iter}$ & s    & $\hat{e}^{imp}$ & $\hat{e}^{iter}$ & s    & $\hat{e}^{imp}$ & $\hat{e}^{iter}$ \\
error mechanism &      &                 &                  &      &                 &                  &      &                 &                  &      &                 &                  &      &                 &                  \\ \hline
add edges (0.1) & 89.1 & 2.0             & 1.9              & 93.3 & 2.4             & 2.2              & 94.2 & 1.8             & 1.8              & 89.0 & 2.0             & 2.0              & 89.7 & 1.9             & 1.7              \\
rm e prop (0.1) & 88.0 & 2.1             & 2.1              & 90.9 & 2.6             & 3.2              & 93.6 & 1.9             & 1.8              & 87.9 & 2.0             & 2.0              & 88.5 & 1.9             & 1.9              \\
rm e unif (0.1) & 88.2 & 2.3             & 2.2              & 91.2 & 2.6             & 2.7              & 93.9 & 1.8             & 2.0              & 88.5 & 2.0             & 2.0              & 88.8 & 2.1             & 1.9              \\
rm nodes (0.1)      & 89.2 & 2.8             & 2.4              & 93.0 & 2.2             & 2.4              & 94.9 & 1.8             & 2.0              & 89.4 & 3.0             & 2.9              & 89.7 & 2.0             & 1.9              \\
add edges (0.3) & 82.1 & 3.0             & 3.0              & 86.1 & 3.7             & 3.5              & 86.6 & 3.4             & 3.1              & 81.9 & 3.5             & 3.4              & 82.9 & 3.0             & 3.1              \\
rm e prop (0.3) & 77.0 & 5.1             & 4.8              & 77.9 & 5.8             & 5.9              & 81.9 & 5.5             & 4.4              & 76.7 & 5.1             & 5.0              & 77.4 & 5.1             & 4.6              \\
rm e unif (0.3) & 79.2 & 3.7             & 3.7              & 80.6 & 4.2             & 3.8              & 84.6 & 3.6             & 4.3              & 79.5 & 3.7             & 3.7              & 80.0 & 3.5             & 3.5              \\
rm nodes (0.3)      & 80.5 & 5.0             & 4.2              & 83.4 & 5.3             & 4.8              & 86.2 & 5.0             & 4.8              & 80.9 & 7.0             & 4.7              & 81.2 & 3.9             & 3.7              \\
\hline\hline
\end{tabular}

\end{table}

It can be seen from \reftable{tab:er95res} that there is a wide range of absolute error values (ranging from 0.017 to .07).
How should these values be interpreted? 
For example, the estimates for the sensitivity of the degree centrality and the PageRank under the influence of the add edges 10\% error mechanism show approximately the same error values (ranging from 0.017 to 0.019).
However, we argue that the estimate for the PageRank works better because the (average) sensitivity of the PageRank is 0.897 while the average sensitivity of the degree centrality is substantially higher (0.942).
Therefore, one has to consider the magnitude of the true sensitivity when interpreting the absolute error of an estimate for the sensitivity.

\subsection{Evaluating estimation results}
\label{sec:EvaluationDescription}
The estimate for the true sensitivity is successful if it is ``close" to the 
true sensitivity value. To determine this ``closeness", we calculate the error (absolute difference between true value 
and estimate) relative to the magnitude of the true sensitivity. On the one hand, if the true sensitivity 
is low, the estimate does not need to be as accurate as if the true sensitivity is high. On the 
other hand, we assume that estimating the sensitivity is more difficult if the true sensitivity is 
relatively low. Moreover, for the evaluation of the ``closeness'' between the true sensitivity~$s$ and an estimate for the true sensitivity $\hat{s}$, we ignore the 
direction of the deviation. Therefore, we define the weighted error of the estimate as
\begin{equation}
\label{eq:errorfunction}
weightederror(s, \hat{s}) := \frac{|s - \hat{s}|}{1 - s} \textrm{ for } s < 1,
\end{equation}
\noindent
where smaller values indicate better 
performance. For the ultimate decision whether the estimate is close enough to the target value, 
we use an indicator function which takes the value one if $weightederror(s, \hat{s})$ is below a given 
threshold value and zero otherwise.

\begin{figure}[htbp]
\includegraphics[width=12cm]{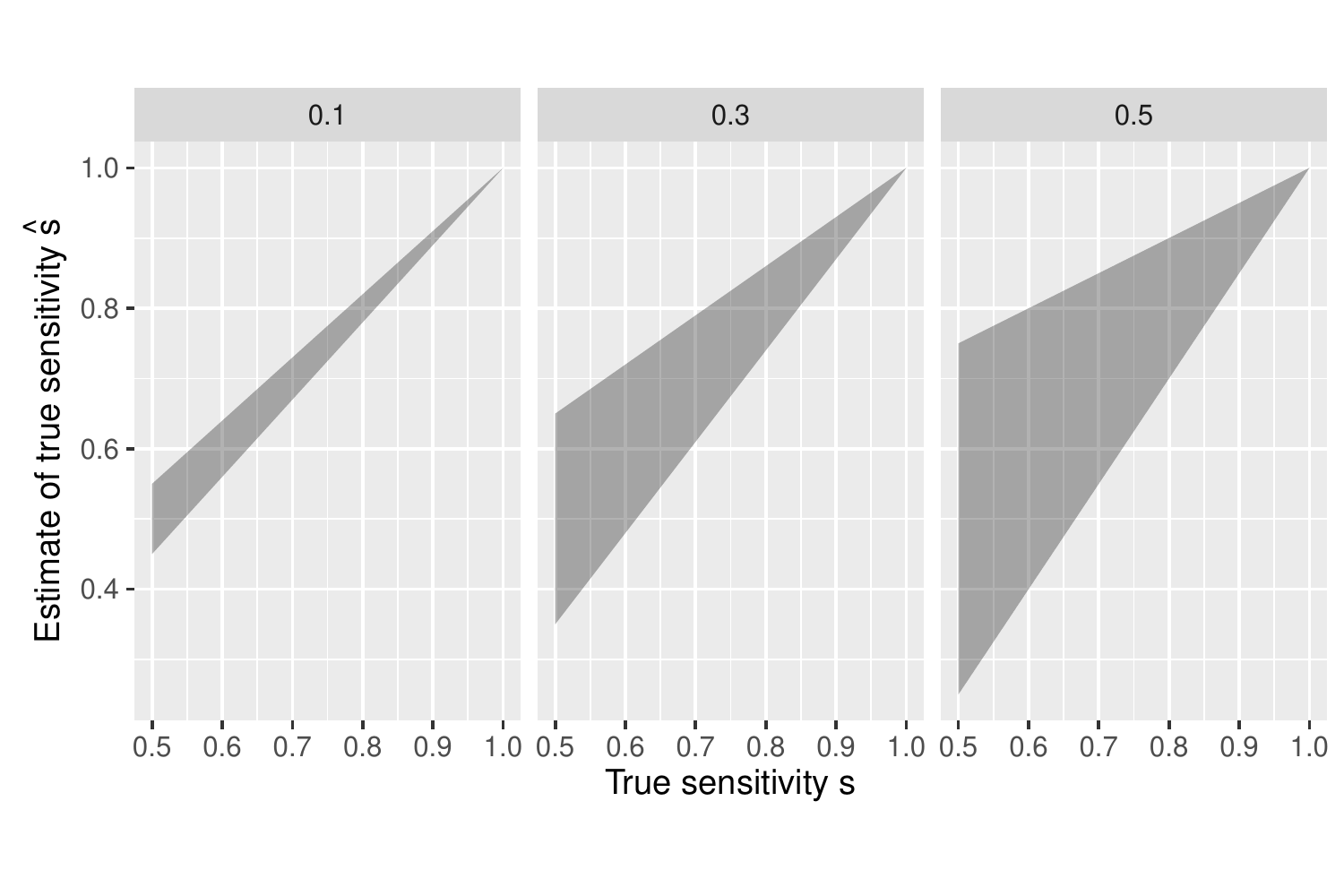}
\caption[Shortened figure caption for the list of illustrations]
{
The dark areas in the panels above show successful combinations of true sensitivity~$s$ and
estimate of the true sensitivity~$\hat{s}$ for three threshold values (from left to right: 0.1, 0.3, and 0.5). 
An estimate of the true sensitivity is only considered successful if the pair 
$(s, \hat{s})$ is within the dark area.

}
\label{fig:errorfunction}
\end{figure}

\reffigure{fig:errorfunction} illustrates the values of this indicator function combined with threshold values of 0.1, 0.3, and 0.5.
The dark areas indicate combinations of true sensitivity and estimate of the true sensitivity that we consider successful with respect to the particular threshold value.
For example, with a threshold value of 0.1, the estimate of the sensitivity has to be very close to the true value, even for low sensitivity values. In this study, we focus on a threshold value of 0.3.
We will use this approach for the remainder of this study to evaluate the performance of the  estimate.

The weighted error combined with a threshold has two important advantages compared to the absolute error. It takes the variation within the experimental runs into account and requires estimates for higher sensitivity values to be more precise than estimates for lower sensitivity values. Applied to the previous example, we get the following success rates (ratio between the number of successful estimates and the total number of estimates):
imputation estimate 0.938, iterative estimate 0.910 in the case of degree  and imputation estimate 0.996, iterative estimate 1.000 in the case of PageRank.
Using the weighted error, we notice that the estimates for the sensitivity of the degree centrality are very good and estimates for the sensitivity of PageRank are remarkable.

\subsection{Results for synthetic graphs}
\label{sec:erResults}

\begin{figure}[htbp]
\includegraphics[width=.9\textwidth]{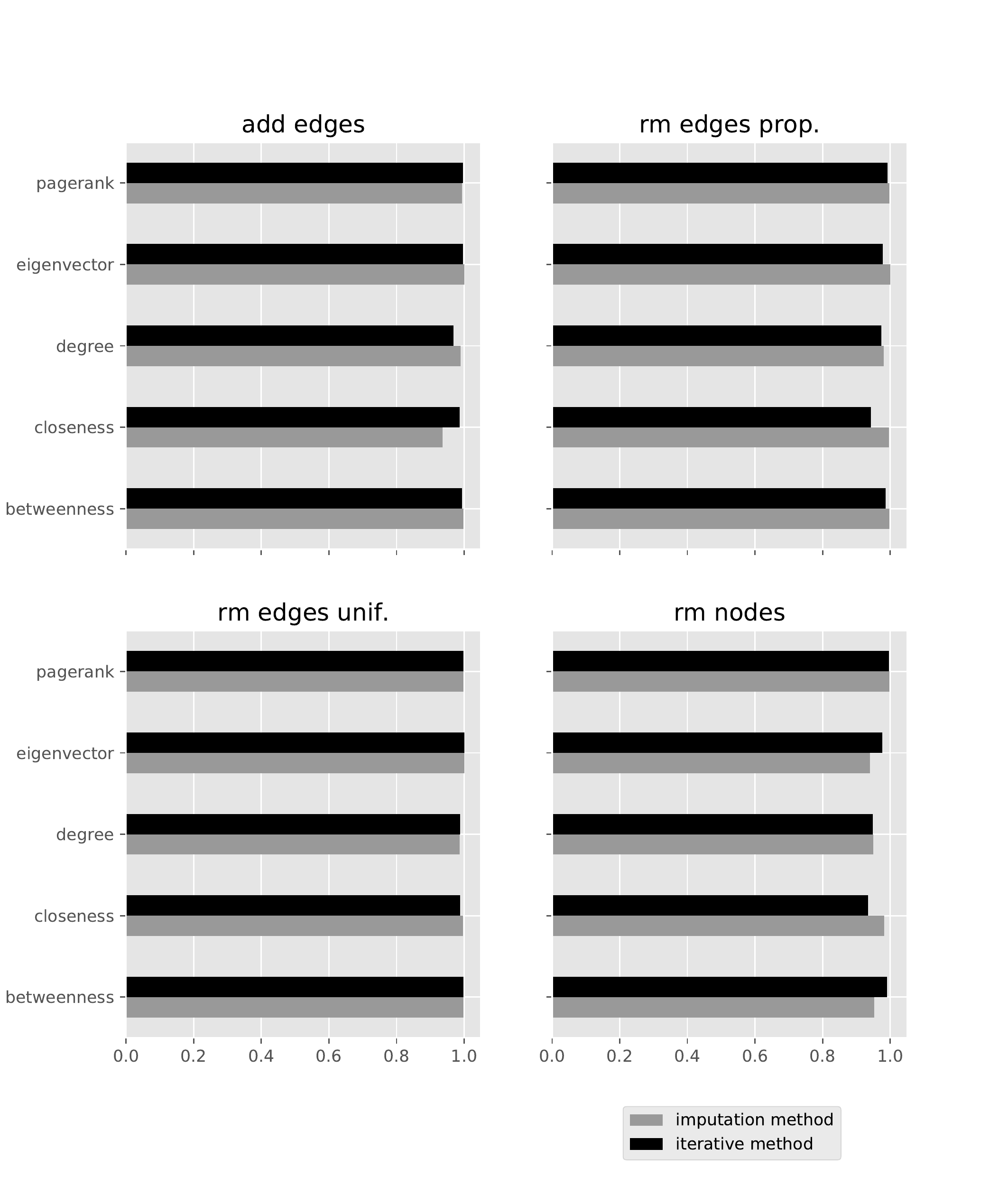}
\caption[Shortened figure caption for the list of illustrations]
{
The success rates for the estimate (ER graphs, 30\% error level) are shown in the figure above. The bar length
indicates the percentage of successful cases among all trials (success rate).  
Grey (black) bars represent the success rates for the imputation (iterative) method.
}
\label{fig:erresultsindicator}
\end{figure}

\reffigure{fig:erresultsindicator} illustrates the performance of the estimates for the experiments on ER graphs.
In general, the performance of the estimate for the sensitivity in the context of ER graphs is remarkable. The success rate, 
i.e. the fraction of cases where the estimate is within the boundaries as defined in \refsection{sec:EvaluationDescription}, is largely 
above 90\%.

The success rates for all 30\% error mechanisms and centrality measures are illustrated in \reffigure{fig:erresultsindicator}.
In most cases, there is no difference between iterative and imputation estimate except for cases that involve the closeness centrality. In those cases, the imputation estimate is better if edges or nodes are missing and the iterative method performs better if there are additional edges. This effect diminishes with increasing intensity.

The success rates for 
betweenness centrality, eigenvector centrality, and PageRank at the same, very high, level, followed by closeness and degree centrality. The 
latter two show slightly  lower (but still high) success rates. The different error mechanisms show similar results.  When comparing the 10\% 
and 30\% error mechanism, we cannot observe that the success rates are lower for the latter. The converse seems to be the case. 
The success rates for cases that involve error mechanisms with 30\% error level are higher than the corresponding values for 10\% error 
mechanisms. At first, this observation seems counter-intuitive. However, we also observe that the sensitivity decreases with increasing 
measurement error.  Since the sensitivity is lower, the interval for valid estimates becomes larger (\refequation{eq:errorfunction}) and
in the case of ER graphs, there is low variation and the success rates become better with increasing intensity.

\begin{figure}[htbp]
\includegraphics[width=.9\textwidth]{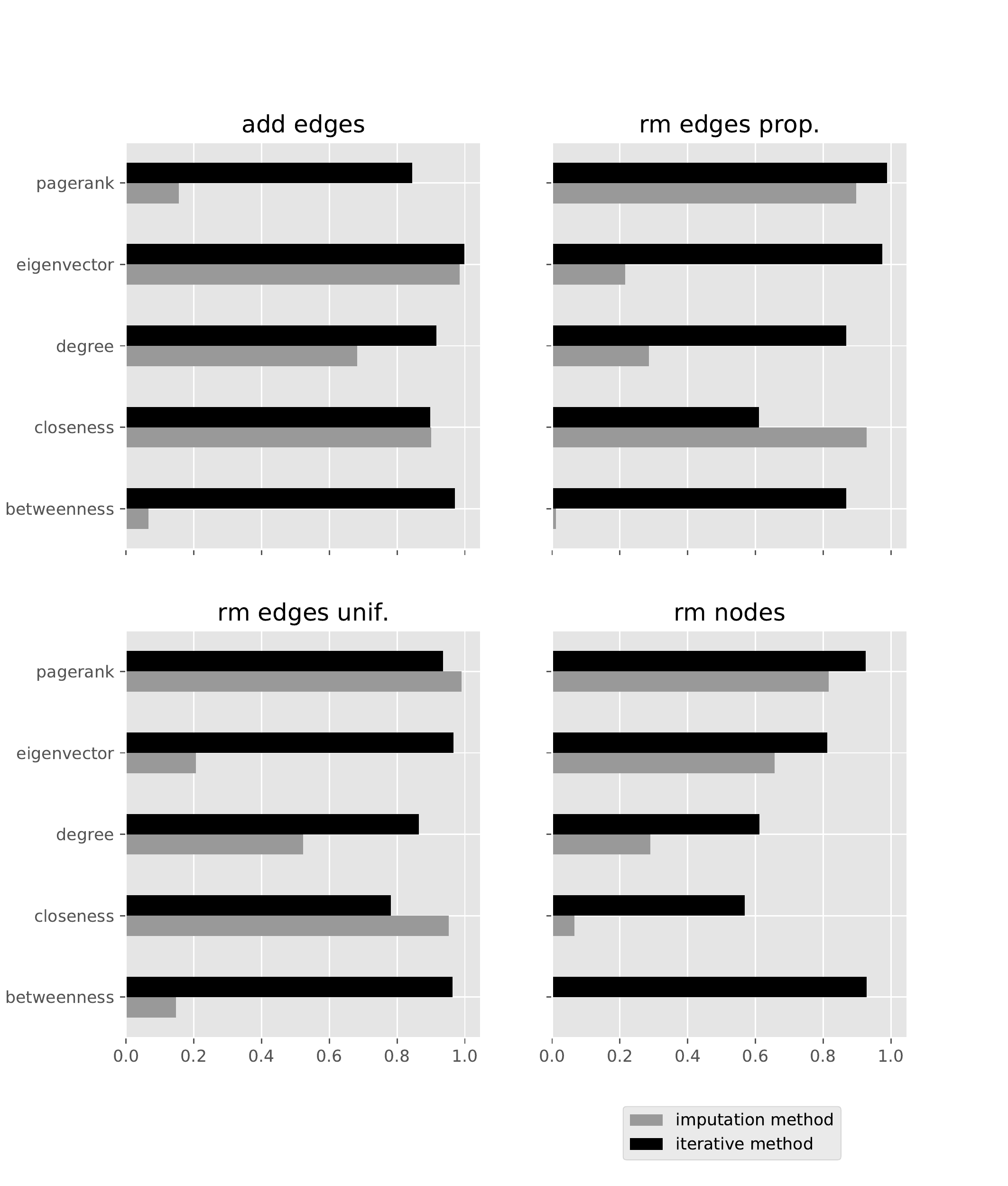}
\caption[Shortened figure caption for the list of illustrations]
{
The success rates for the estimate (BA graphs, 30\% error level) are shown in the figure above. The bar length
indicates the percentage of successful cases among all trials (success rate).  
Grey (black) bars represent the success rates for the imputation (iterative) method. 
}
\label{fig:baresultsindicator}
\end{figure}

We also perform the experiment described in \refsection{sec:ERexperimentSetup} with one difference in the first step: instead of an ER graph, we generate a BA graph \cite{Barabasi1999a}, 100 nodes, parameter m = 11, undirected).
Results for this experiment for cases with 30\% error level are shown in \reffigure{fig:baresultsindicator}.
In general, the imputation method performs worse than the iterative method. The performance of the imputation method is particularly bad in cases where the betweenness centrality has to be estimated.

In contrast, the iterative method shows high success rates in most of the cases. In cases that involve the degree or closeness centrality, the performance is usually worse that in the remaining cases. There is little difference between the four error mechanisms.
The results for the 10\% error mechanism (Appendix~\ref{sec:AdditionalResults}) are similar. However, in some cases, we observe lower success rates than in for the corresponding 30\% error mechanism.

\section{Application to real-world networks}
\label{sec:applicationToRw}
Here, we apply our methods from Section~\ref{sec:MethodDescription} to real-world networks in 
order to investigate the suitability of these methods for practical application.
We use four networks from different domains and thus different structural properties to get an impression how these 
methods perform on real data. Descriptive statistics for these networks
are listed in 
\reftable{tab:EmpiricalNetworksStatistics}.

\begin{table}[h]
\centering
\caption{Statistics of real-world networks. If the original network is not connected, we only consider the largest connected component.}
\setlength{\tabcolsep}{0.2em} 
\label{tab:EmpiricalNetworksStatistics}

\begin{tabular}{lrrrrrl}
\hline\hline
Network     & Nodes & Edges  & Clustering & Density & Diameter & Source             \\ \hline
Dolphins    & 62    & 159    & 0.3029     & 0.0841  & 8        & \cite{Lusseau2003} \\
Jazz        & 198   & 2,742  & 0.6334     & 0.1406  & 6        & \cite{Gleiser2003} \\
Protein     & 1,458 & 1,948  & 0.1403     & 0.0018  & 19       & \cite{Jeong2001}   \\
Hamsterster & 1,788 & 12,476 & 0.1655     & 0.0078  & 14       & \cite{Kunegis2013} \\ 
\hline\hline
\end{tabular}

\end{table}

\label{sec:empiricalsetup}
We use our proposed methods to estimate the sensitivity of five centrality measures under the influence of four error mechanisms. For every error mechanism, we consider two cases, a moderate scenario of 10\% error level and a more intense scenario with 30\% error level. 
 For every combination of network, centrality measure, and error mechanism, the experimental setup is as follows:
 \begin{enumerate}
\item
Due to the very nature of the hidden networks, we cannot access them.  Hence, for the sake of our experiments, we treat the real-world network as the error-free  hidden network $H$.
 (This is a common approach used in existing studies about the sensitivity of centrality measures.)
 \item
To simulate erroneous data collection, we choose a graph from \errG{H} and denote it by $O$. This graph represents the observed network which is 
affected by measurement errors. For evaluation purposes, the true sensitivity  $\rho_{c}(H, O)$ is 
calculated and denoted by $s$.
\item
Based on the observed network $O$, two estimates for the true sensitivity are calculated.
The imputation estimate ($\hat{\rho}_c^{imp}(O, H) $) is denoted by $\hat{s}^{imp}$,
the estimate calculated according to the iterative method ($\hat{\rho}_c^{iter}(O, H) $) is denoted by $\hat{s}^{iter}$.
\end{enumerate}

\noindent
For every combination, we perform this experiment 500 times. To evaluate the results, 
we use the procedures described in Section \ref{sec:EvaluationDescription}.

\subsection{Results for real-world networks}
In this section, we study how our methods for estimating the sensitivity of centrality measures perform on real-world networks.
Regarding the iterative estimates, we observe a fair amount of cases with high success rate. However, the results for empirical networks are more heterogeneous than the 
results for Erdős–Rényi and Barabási–Albert networks (\refsection{sec:erResults}). But since the real-world networks are more complex than 
graphs generated by these procedures,
we expected that our estimation methods would not work as well for real-world networks compared to synthetic networks.

\begin{figure}[htbp]
\includegraphics[width=1\textwidth]{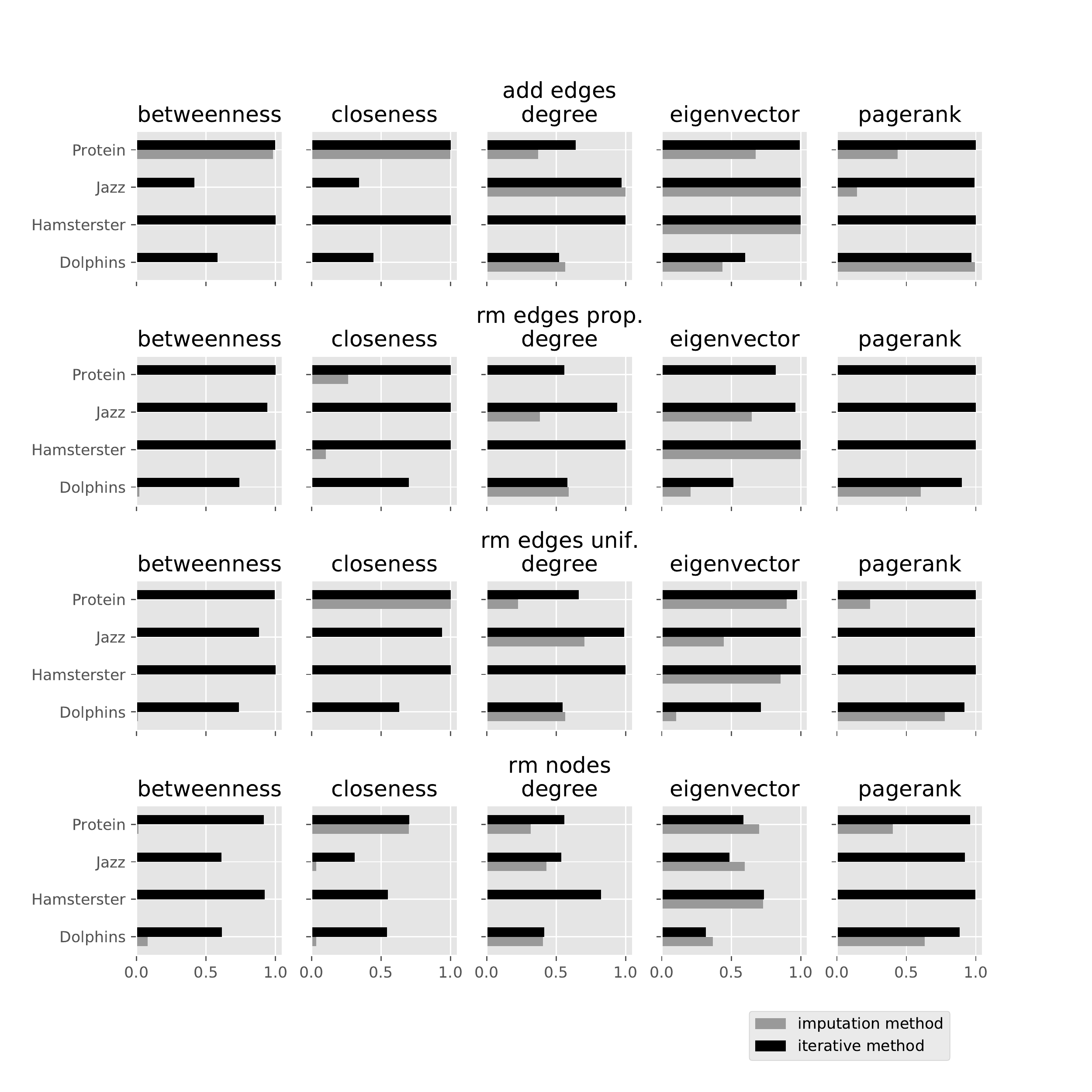}
\caption[Shortened figure caption for the list of illustrations]
{
The success rates for the estimate (real-world networks, 10\% error level) are shown in the figure above. The bar length
indicates the percentage of successful cases among all trials (success rate).  
Grey (black) bars represent the success rates for the imputation (iterative) method.
}
\label{fig:rwresultindicator1}
\end{figure}

\begin{figure}[htbp]
\includegraphics[width=1\textwidth]{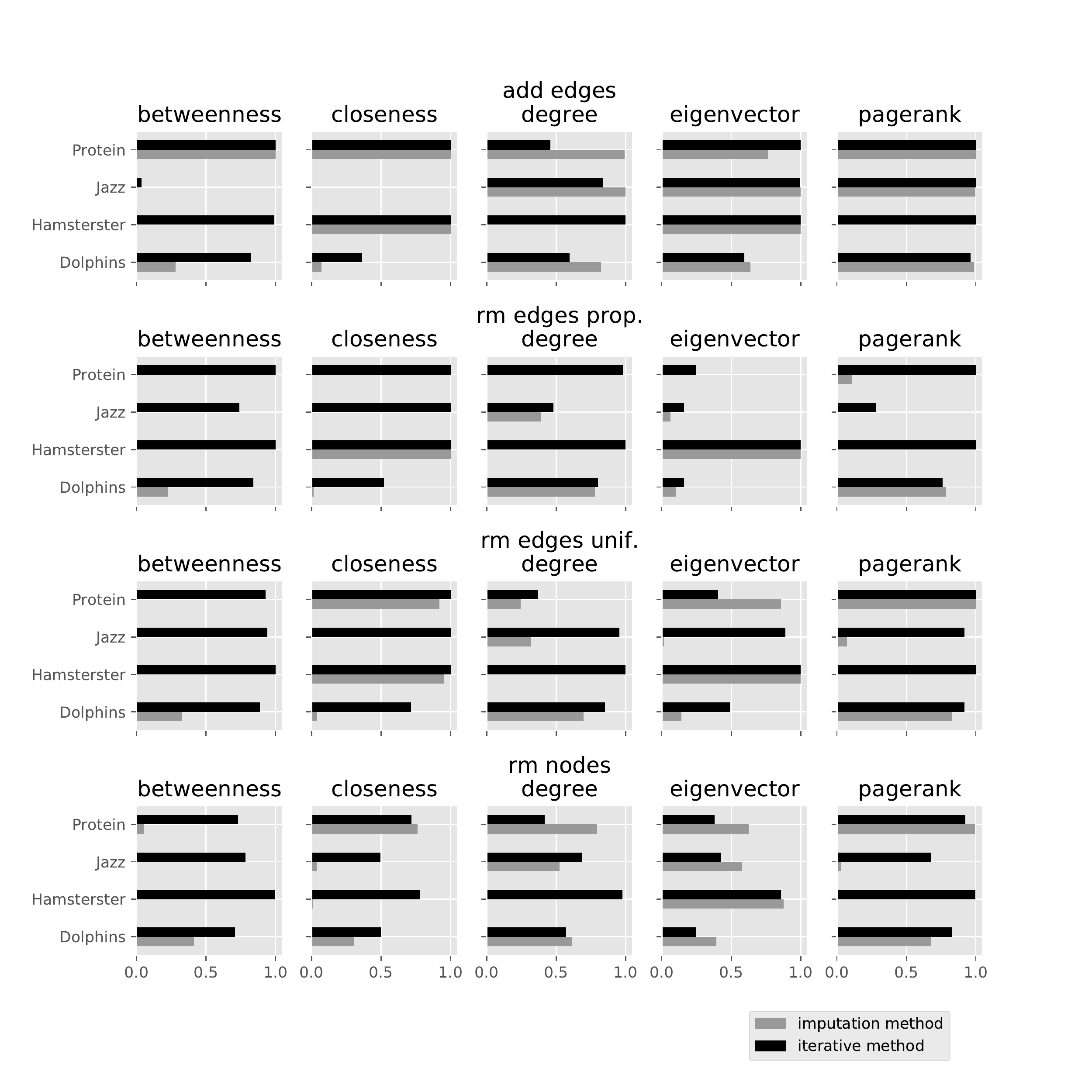}
\caption[Shortened figure caption for the list of illustrations]
{
The success rates for the estimate (real-world networks, 30\% error level) are shown in the figure above. The bar length
indicates the percentage of successful cases among all trials (success rate).  
Grey (black) bars represent the success rates for the imputation (iterative) method.
}
\label{fig:rwresultindicator3}
\end{figure}

The results for the estimate of the sensitivity of centrality measures in  real-world networks are shown in 
\reffigure{fig:rwresultindicator1} and~\ref{fig:rwresultindicator3}.
First, we focus on the results for error mechanisms with 10\% error level (\reffigure{fig:rwresultindicator1}).
Comparing both estimation methods, it turns out that the iterative method is at least as good as the imputation method, except for a few cases where the former is slightly better. Hence, we first focus on the estimates of the iterative method.
The iterative estimate for PageRank works in virtually all cases, regardless of the specific network or error mechanism.  Among the four 
networks, the success rates for the Dolphins network are usually the lowest. It is reasonable to assume that this effect is due to the small 
size of the Dolphin network (62 nodes, 159 edges). If we focus our discussion on the three larger networks, we observe high success rates 
for the estimates of the sensitivity of closeness, betweenness, and eigenvector centrality if edges are missing uniformly or proportional. 
For these networks, the estimates for the sensitivity of the eigenvector centrality also work if the measurement errors lead to too many 
edges in the observed network.
The results suggest, that the error mechanism missing nodes is most difficult  for the estimate.
There is no strong relationship between sensitivity and the success rate, higher sensitivity values are not easier to estimate. The average sensitivity values for all real-world networks can be found in Appendix~\ref{sec:AdditionalResults}.

Comparing \reffigure{fig:rwresultindicator1} and~\ref{fig:rwresultindicator3} shows that the success rates for the iterative method become lower with an increasing 
level of error. It is more difficult to estimate the sensitivity for higher levels of measurement errors.
The estimate for the sensitivity of PageRank still works  for error mechanisms additional edges and missing edges.
If we focus on the three largest networks, the cases involving betweenness and closeness centrality show good success rates if edges are 
missing (uniformly and proportional). Cases involving the eigenvector centrality show good success rates if edges are missing uniformly and 
for the error mechanism additional edges.

We observe cases with a subpar performance for 10\% error level where the success rates continue to decrease. For example, the Jazz 
network with additional edges in combination with the closeness and betweenness centrality. 
There are few cases that show good performance for 10\% error level but work barely for 30\% error level (e.g., if edges are missing proportionally in the Jazz network and we try to estimate the sensitivity of eigenvector centrality or PageRank).

The results for the imputation estimates of the sensitivity are rather different. There is no centrality measure or error mechanism where this method shows high success rates for all four networks. In some cases, the imputation estimate for error levels of 30\% is better than the imputation estimate for the corresponding 10\% case (e.g., PageRank and additional edges). Most of these situations occur when the error mechanism is additional edges.

\section{Discussion and conclusion}
\label{sec:discussion}

Errors in network data are a ubiquitous problem in network analysis and previous studies have shown that these errors can have a severe impact on the reliability of centrality measures.
Most studies that use centrality measures, however, rarely discuss the ramifications that measurement errors have on their analyses. Usually, these studies mention that the observed data might contain errors, but analyses are performed as if the data is error-free.
Even though the reliability of centrality measures has been studied extensively,
there is no technique that  allows researchers to assess the reliability of centrality measures in the case of imperfect observed data.

In the first part of this study, we introduced concepts to describe such a technique. We defined an easy-to-interpret metric, the sensitivity, to measure the reliability of centrality measures.
Additionally, we presented the concept of error mechanisms, which model measurement errors as random graphs.
We applied these concepts to ER graphs and the results are consistent with previous research \cite{Borgatti2006a}.

In the main part of this study, we proposed two methods (``imputation method" and ``iterative method") that allow the researcher to estimate the sensitivity of the error-free (hidden) network, given the observed network and some assumptions about the measurement error.

Our experiments showed that both methods performed very well on ER graphs. In the case of BA graphs and real-world networks, the imputation method rarely worked and should therefore not be used.
These findings extend those of Huisman (2009) \cite{Huisman2009}, confirming that imputation methods are only useful in a few specific situations.
Surprisingly, the method that is easier to calculate yielded better results.
We could  identify cases where the iterative method showed remarkable performance. It worked especially well for the PageRank for all error mechanisms with 10\% error level. If the error level increased to 30\%, the iterative method still showed good performance if edges were missing uniformly at random or if there were spurious edges.
If 10\% of the edges were missing uniformly at random or proportional and the network was not too small, the iterative method performed well for all centrality measures except for the degree centrality. The sensitivity values for the degree centrality were, however, relatively high.

Our results provide compelling evidence that the iterative method is, in principle, a suitable technique for the estimation of the sensitivity of centrality measures. Hence, it is a promising first step that helps researchers to assess the impact of measurement errors on their observed network data.
Although the iterative method works in well in many cases, there are limitations. There is a need to clarify the conditions under which the self-similarity assumption does not hold true and thus identify, based on the observed network, the cases where the iterative method should not be used.
Another important question for future studies is to determine more suitable imputation mechanisms and thus improving the performance of the imputation method.

\clearpage
\appendix
\section{Additional results}
\label{sec:AdditionalResults}

\begin{figure}[htbp]
\includegraphics[width=.9\textwidth]{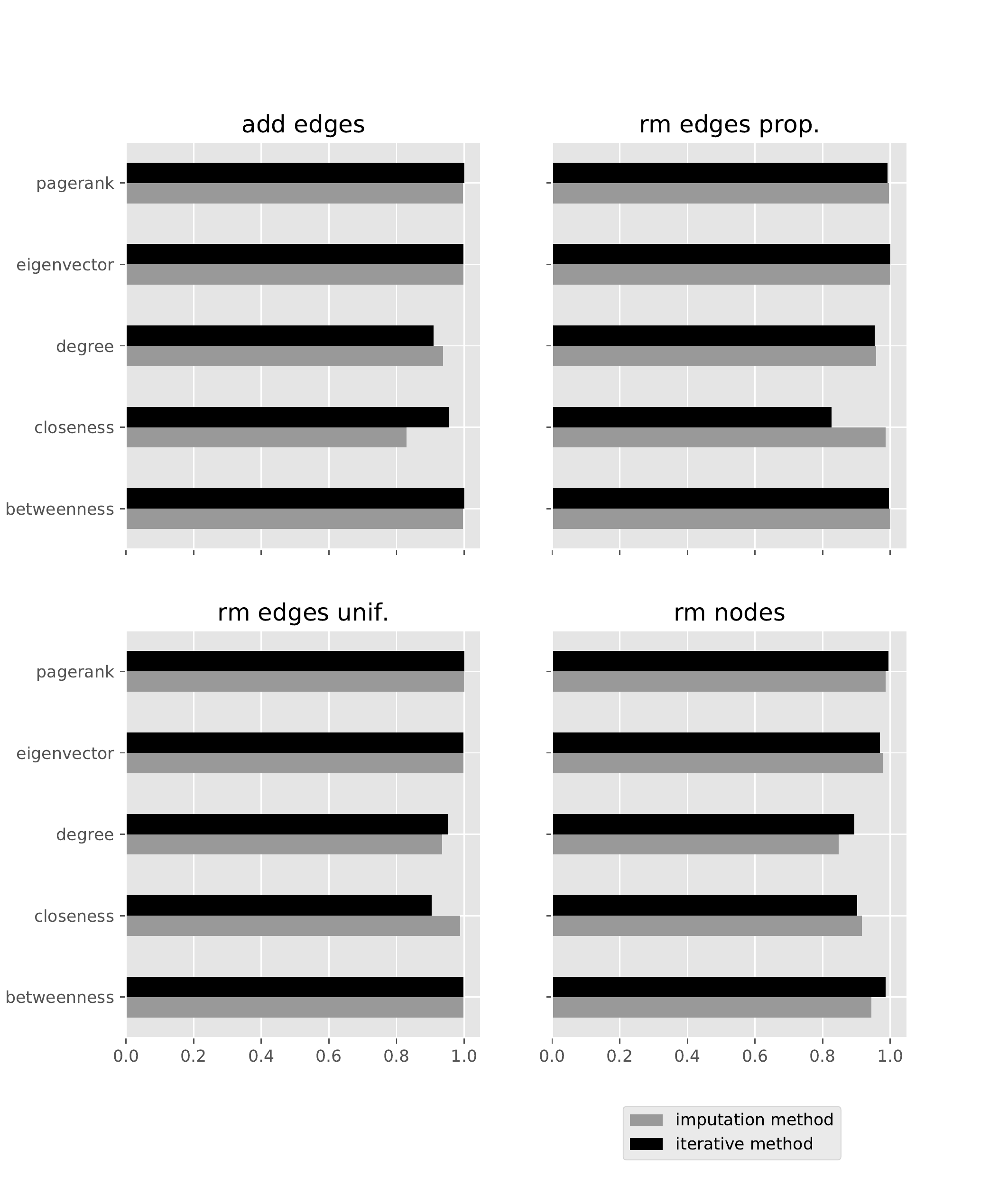}
\caption[Shortened figure caption for the list of illustrations]
{
The success rates for the estimate (ER graphs, 10\% error level) are shown in the figure above. The bar length
indicates the percentage of successful cases among all trials (success rate).  
Grey (black) bars represent the success rates for the imputation (iterative) method.
}
\end{figure}

\begin{figure}[htbp]
\includegraphics[width=.9\textwidth]{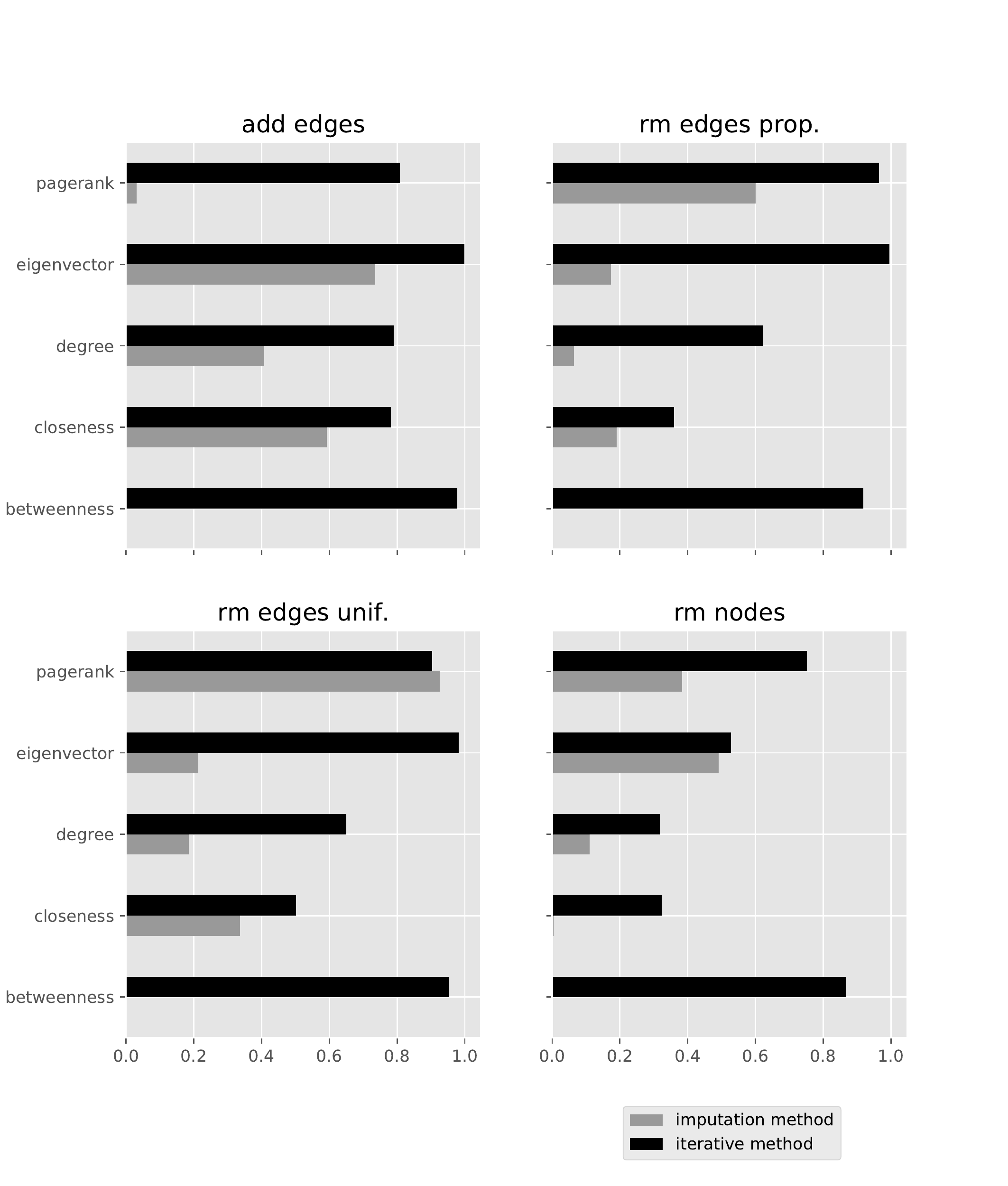}
\caption[Shortened figure caption for the list of illustrations]
{
The success rates for the estimate (BA graphs, 10\% error level) are shown in the figure above. The bar length
indicates the percentage of successful cases among all trials (success rate).  
Grey (black) bars represent the success rates for the imputation (iterative) method.
}
\end{figure}

\begin{table}[h]
\centering
\caption{The average sensitivity values for the real-world networks are listed in the table below.
(Values are multiplied by 100 for better readability.) }
\label{tab:SensitivityEmpiricalAppendix}
\setlength{\tabcolsep}{0.08em} 

\begin{tabular}{llrrrrrrrrrr}
\hline\hline
\multicolumn{2}{r}{Centrality measure} & bc   &      & cc   &      & dc   &      & ec   &      & pr   &      \\
\multicolumn{2}{r}{Level of error}     & 0.1  & 0.3  & 0.1  & 0.3  & 0.1  & 0.3  & 0.1  & 0.3  & 0.1  & 0.3  \\ \cline{1-2}
Error mechanism        & Network       &      &      &      &      &      &      &      &      &      &      \\ \hline
Add edges              & Dolphins      & 80.8 & 76.2 & 82.1 & 76.3 & 98.5 & 94.2 & 89.3 & 81.5 & 93.0 & 87.6 \\
                       & Hamsterster   & 84.5 & 81.3 & 94.4 & 90.5 & 97.7 & 94.1 & 96.4 & 93.3 & 91.7 & 87.7 \\
                       & Jazz          & 82.3 & 79.5 & 90.4 & 86.5 & 98.2 & 95.9 & 97.0 & 94.1 & 95.0 & 92.4 \\
                       & Protein       & 92.3 & 86.1 & 90.8 & 83.0 & 99.0 & 95.1 & 91.1 & 83.7 & 89.6 & 81.6 \\
Remove edges (prop.)   & Dolphins      & 91.9 & 83.3 & 93.1 & 85.0 & 98.5 & 92.2 & 92.6 & 76.6 & 93.7 & 85.7 \\
                       & Hamsterster   & 97.1 & 93.4 & 95.7 & 89.5 & 99.5 & 97.6 & 96.3 & 90.7 & 97.6 & 94.6 \\
                       & Jazz          & 96.0 & 91.7 & 95.5 & 90.7 & 98.5 & 95.3 & 97.2 & 94.1 & 97.0 & 93.3 \\
                       & Protein       & 95.4 & 89.7 & 85.0 & 67.2 & 99.6 & 96.9 & 80.2 & 63.0 & 93.5 & 85.6 \\
Remove edges (unif.)   & Dolphins      & 91.4 & 82.9 & 93.2 & 85.3 & 98.4 & 93.3 & 93.7 & 85.5 & 93.4 & 86.1 \\
                       & Hamsterster   & 96.0 & 91.2 & 96.6 & 91.6 & 99.3 & 97.0 & 96.8 & 92.6 & 96.3 & 92.1 \\
                       & Jazz          & 95.1 & 89.6 & 96.2 & 91.5 & 98.4 & 95.8 & 97.5 & 94.9 & 96.5 & 93.1 \\
                       & Protein       & 95.5 & 90.7 & 88.8 & 75.8 & 99.3 & 95.5 & 87.9 & 76.3 & 92.2 & 82.9 \\
Remove nodes           & Dolphins      & 90.9 & 82.8 & 92.2 & 84.5 & 98.4 & 93.8 & 90.1 & 80.7 & 93.5 & 87.1 \\
                       & Hamsterster   & 96.4 & 91.7 & 96.7 & 91.9 & 99.3 & 97.0 & 96.7 & 92.4 & 96.5 & 92.4 \\
                       & Jazz          & 95.5 & 90.5 & 96.8 & 92.8 & 98.7 & 96.4 & 97.0 & 93.5 & 97.3 & 94.4 \\
                       & Protein       & 95.6 & 91.2 & 88.3 & 75.0 & 99.4 & 95.4 & 87.0 & 75.3 & 92.2 & 83.1 \\ \hline\hline
\end{tabular}
\end{table}

\clearpage
\bibliography{/home/crs/Documents/MendeleyBibtex/library.bib}
\end{document}